\documentclass[longbibliography,reprint,2,amsmath,amssymb,aps]{revtex4-1}

\usepackage{graphicx}
\usepackage{dcolumn}
\usepackage{bm}
\usepackage{natbib}
\usepackage[table]{xcolor}
\usepackage{tabularx}
\usepackage{array}
\usepackage{colortbl}
\usepackage{appendix}
\usepackage{amsmath}
\usepackage{makecell}

\newcolumntype{L}[1]{>{\raggedright\arraybackslash}p{#1}}

\begin{document}
\preprint{APS/123-QE}

\title{Time-Frequency Grid States for Reconstruction and Correction of Channel-Induced Distortion in Entangled Photons}

\author{Siang-Yun Liu $^{1}$, Bo-Ren Huang $^{2}$, Zhi-Xuan Tseng $^{2}$, Yen-Hung Chen $^{2,3,4,\dagger}$, Pin-Ju Tsai $^{1,2,3}$}
\email{tpinju@ncu.edu.tw}
\email{$^{\dagger}$yhchen@dop.ncu.edu.tw}

\affiliation{$^{1}$Department of Physics, National Central University, Taoyuan City 320317, Taiwan
}
\affiliation{$^{2}$Department of Optics and Photonics, National Central University, Taoyuan City 320317, Taiwan
}
\affiliation{$^{3}$Quantum Technology Center, National Central University, Taoyuan City 320317, Taiwan
}
\affiliation{$^{4}$Center for Astronautical Physics and Engineering, National Central University, Taoyuan City 320317, Taiwan
}

\begin{abstract}

Characterization of time-frequency (TF) quantum states requires reliable reconstruction of their TF distributions. However, imperfect transmission or measurement channels can distort reconstructed joint spectral intensities (JSIs), especially when the underlying perturbation mechanism is unknown. Here, we experimentally demonstrate a reconstruction and correction framework that uses a TF grid state as an intrinsic frequency-domain reference. By analyzing the displacement of the grid points, a Gaussian process regression model is employed to reconstruct a correction mapping for the nonlinear coordinate deformation without assuming a prior physical model of the distortion. The learned mapping reduces the residual coordinate deviation of the TF grid state by approximately a factor of 11 and, when applied to an independent frequency-entangled test state, improves the Gaussian-shape fidelity from 76.2\% to 90.0\%. These results establish TF grid states as practical metrological resources for diagnosing and correcting distortions in TF quantum systems, providing a pathway toward distortion-resilient quantum communication and high-dimensional quantum information processing.

\end{abstract}
\maketitle
\section{Introduction} \label{secI}

\begin{figure*}[ht]
\centering
\includegraphics[width=0.88\textwidth]{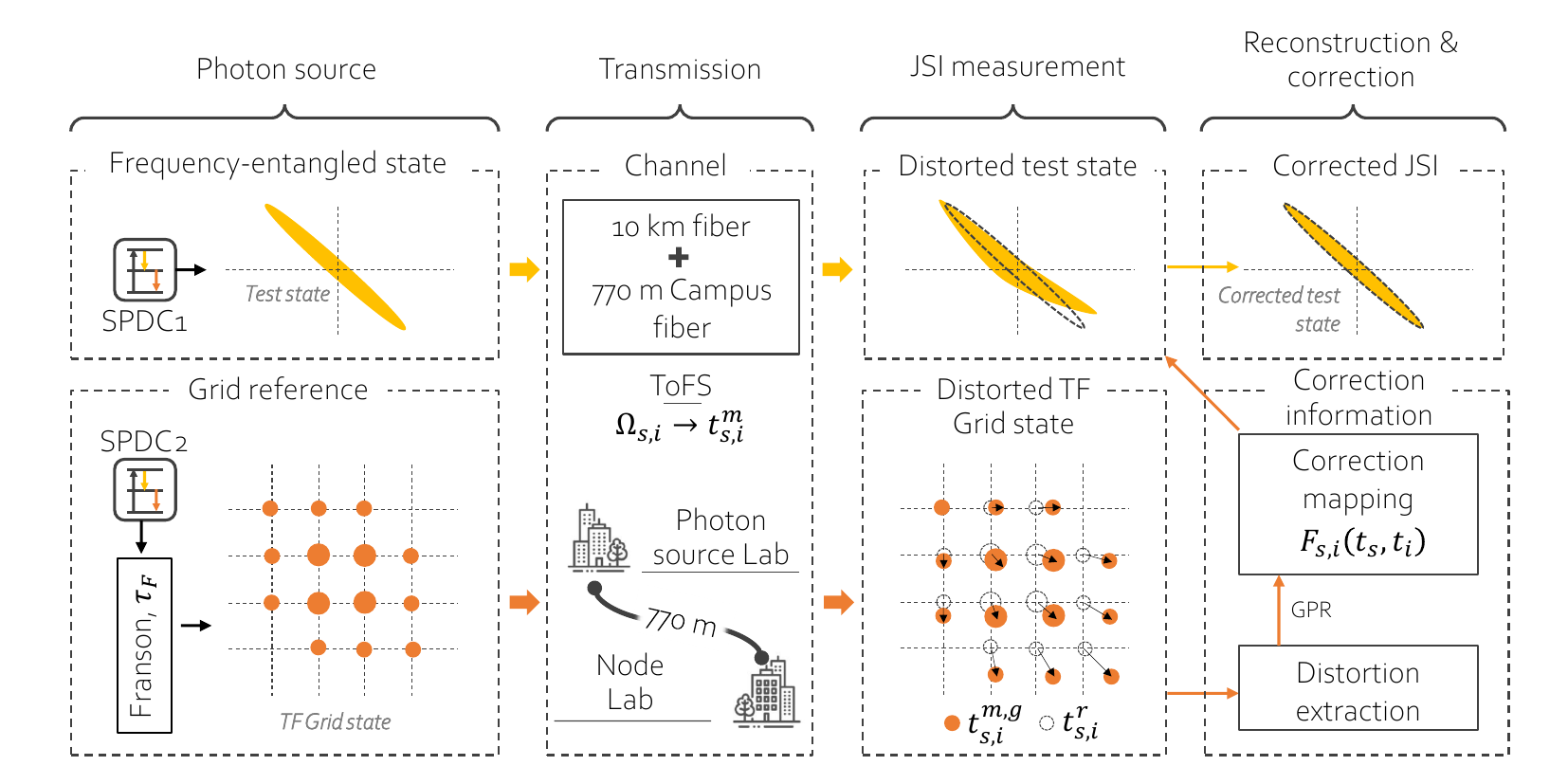}
\caption{End-to-end flow chart from photon-source preparation to distortion reconstruction and JSI correction. $\Omega_{s,i}$ represent the frequencies of the signal and idler photons, respectively. $t^m_{s,i}$ denote the measured arrival times after frequency-to-time mapping through the ToFS. $t_{s,i}^{m,g}$ and $t_{s,i}^{r}$ denote the measured and reference grid-point coordinates, respectively. $F_{s,i}(t_s,t_i)$ represents the learned correction mapping.}
\label{Flow}
\end{figure*}

Time–frequency (TF) degrees of freedom provide a high-dimensional platform \cite{fabre2022time} for photonic quantum information and have been widely explored in applications such as quantum key distribution (QKD) \cite{mower2013high,liu2019energy,chang2024large,liu2024high}, quantum computation (QC) \cite{chang2026gkp}, quantum error correction (QEC) \cite{fabre2023teleportation,descamps2024gottesman,yoon2026hardware}, and frequency-domain quantum information processing \cite{brecht2015photon,gianani2020measuring,chang2025recent,yu2024time,lu2023frequency,fabre2025photonic,pousset2026time}. Accurate characterization of TF quantum states, including their joint spectral amplitude (JSA) and joint spectral intensity (JSI), is therefore essential for both fundamental studies and practical implementations.


Various spectral measurement techniques have been developed for this purpose, including single-photon spectrometers based on monochromator or grating scanning \cite{maclean2018direct,maclean2019reconstructing} and time-of-flight spectrometers (ToFSs) based on frequency-to-time mapping through dispersive media. In particular, ToFS has become a widely used platform for TF-state characterization and frequency-encoded quantum information systems \cite{graffitti2020direct,jin2024spectrally,merkouche2022heralding,merkouche2022spectrally,mower2013high,chang2024large,liu2019energy}.

However, the reconstruction of TF quantum states ultimately relies on an accurate coordinate mapping between photon frequencies and experimentally accessible measurement variables. In realistic measurement systems or transmission channels, this mapping can be distorted by instrumental imperfections, environmental fluctuations, or nonideal dispersive responses. Such distortions directly deform the reconstructed TF distributions and may obscure the underlying spectral correlations of the quantum state. Therefore, a reference structure capable of diagnosing such coordinate deformation is highly desirable.

To mitigate measurement errors, calibration is typically performed using external reference signals or models based on independently measured dispersion parameters \cite{davis2017pulsed}. However, in realistic environments, distortions often arise from the combined influence of multiple mechanisms, making them difficult to identify or accurately model. Consequently, reconstructing and correcting such distortions without prior knowledge of their physical origin remains a significant challenge for TF quantum-state characterization.

To address this challenge, TF grid states \cite{jin2024spectrally,fabre2020generation} offer a unique opportunity. A TF grid state is characterized by a periodic lattice structure in the joint spectrum, which naturally provides a set of reference points in the frequency domain. When a TF grid state propagates through an unknown channel or measurement apparatus, channel-induced coordinate deformation manifests itself as systematic displacement of the grid points. By analyzing these displacements, the underlying coordinate deformation can be inferred directly from the measured TF grid state, enabling the grid state to serve as an intrinsic frequency-domain coordinate reference.

In this work, we experimentally demonstrate the use of a TF grid state as a reference resource for reconstructing and correcting channel-induced coordinate deformation in ToFS-reconstructed TF distributions. Unlike conventional single-frequency or single-photon calibration, the TF grid state probes the same two-photon joint measurement space as the target state and therefore provides a direct two-dimensional coordinate reference for correcting joint TF distributions. To the best of our knowledge, the use of TF grid states for this purpose has not been experimentally demonstrated.

Figure \ref{Flow} illustrates the core concept and overall workflow of the proposed framework. Two independent photon-pair sources are prepared: a TF grid state generated by a Franson interferometer is employed as the reference state, while an independent frequency-entangled state serves as the test state. Both states are characterized using the same non-ideal fiber-based ToFS measurement channel, where photon frequencies are mapped onto arrival times through a dispersive fiber link. Imperfections in this frequency-to-time mapping introduce unknown distortions into the reconstructed TF distributions.

By analyzing the displacement of the measured grid points, we reconstruct the underlying coordinate deformation without assuming any prior knowledge of its physical origin. A Gaussian process regression (GPR) model \cite{williams2006gaussian} is employed to learn a correction mapping from the distorted measured coordinates to the ideal reference coordinates. The learned mapping reduces the residual coordinate deviation of the TF grid state by approximately a factor of 11 and, when applied to the independent frequency-entangled test state, improves the Gaussian-shape fidelity from 76.2\% to 90.0\%. These results demonstrate that TF grid states can serve not only as engineered quantum states, but also as practical metrological resources for diagnosing and correcting distortions in reconstructed TF distributions.

The paper is organized as follows. Section \ref{secII} establishes the theoretical framework of the TF grid state and describes the formation mechanism of the TF grid structure in a polarization-based Franson interferometer. Section \ref{secIII} introduces the experimental setup and measurement methodology, including the polarization-based Franson interferometer and the time-of-flight spectrometer (ToFS) system. Section \ref{secIV A} presents the experimentally measured JSIs of the TF grid state and the frequency-entangled test state prior to correction. Based on the TF grid-state measurements, Section \ref{secIV B} identifies the channel-induced coordinate deformation and reconstructs the corresponding correction mapping. The reconstructed correction mapping is then employed in Section \ref{secIV C} to correct the TF grid state. Section \ref{secIV D} further demonstrates the correction of an independent frequency-entangled test state using the reconstructed mapping developed in this work. Finally, Section \ref{secV} summarizes the main findings of this work and discusses potential future applications. Additional theoretical derivations and experimental demonstration are provided in the Supplementary Materials. Additionally, the Appendix provides a list of symbols.

\section{Theoretical model} \label{secII}

In this section, we derive the JSI of the output state generated by the polarization-based Franson interferometer and show how a TF grid structure naturally emerges in the joint spectrum.

\begin{figure}[t]
\centering
\includegraphics[width=0.38\textwidth]{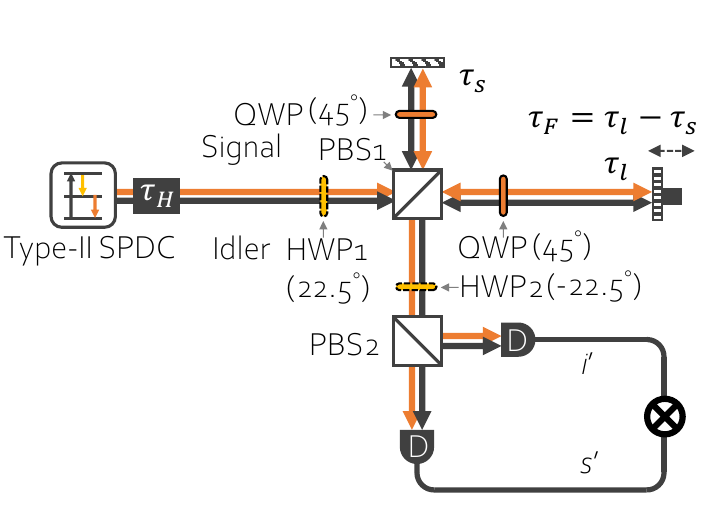}
\caption{Schematic diagram of the polarization-based Franson interferometer.  PBS: polarizing beam splitter; HWP: half-wave plate; QWP: quarter-wave plate; D: single-photon detector; CC: coincidence counting module. The orange and black lines indicate the optical paths of the idler and signal photons, respectively.}
\label{setup_th}
\end{figure}

Figure \ref{setup_th} illustrates the schematic diagram of the setup. Here, we start from the type-II SPDC source that generates a pair of photons exhibiting time-frequency entanglement, whose quantum state can be written as
\begin{equation}
    |\Psi_{\text{SPDC}}\rangle=\int\int d\omega_sd\omega_i f(\omega_s,\omega_i)\hat{a}_H^{\dagger}(\omega_s)\hat{a}_V^{\dagger}(\omega_i)e^{i\omega_s\tau_H}\left|0\right\rangle,
\label{SPDC_QS}
\end{equation}
where $\hat{a}_{H,V}^{\dagger}(\omega)$ represents the creation operator for photons with horizontal (H) and vertical (V) polarizations at frequency $\omega$, and $\omega_{s,i}$ denotes the frequencies of the signal and idler photons, respectively. $f(\omega_s, \omega_i)$ is the JSA of the photon pair. $\tau_H$ is the time delay between signal and idler photons due to their different refractive index in the crystal.

\begin{figure}[t]
\centering
\includegraphics[width=0.48\textwidth]{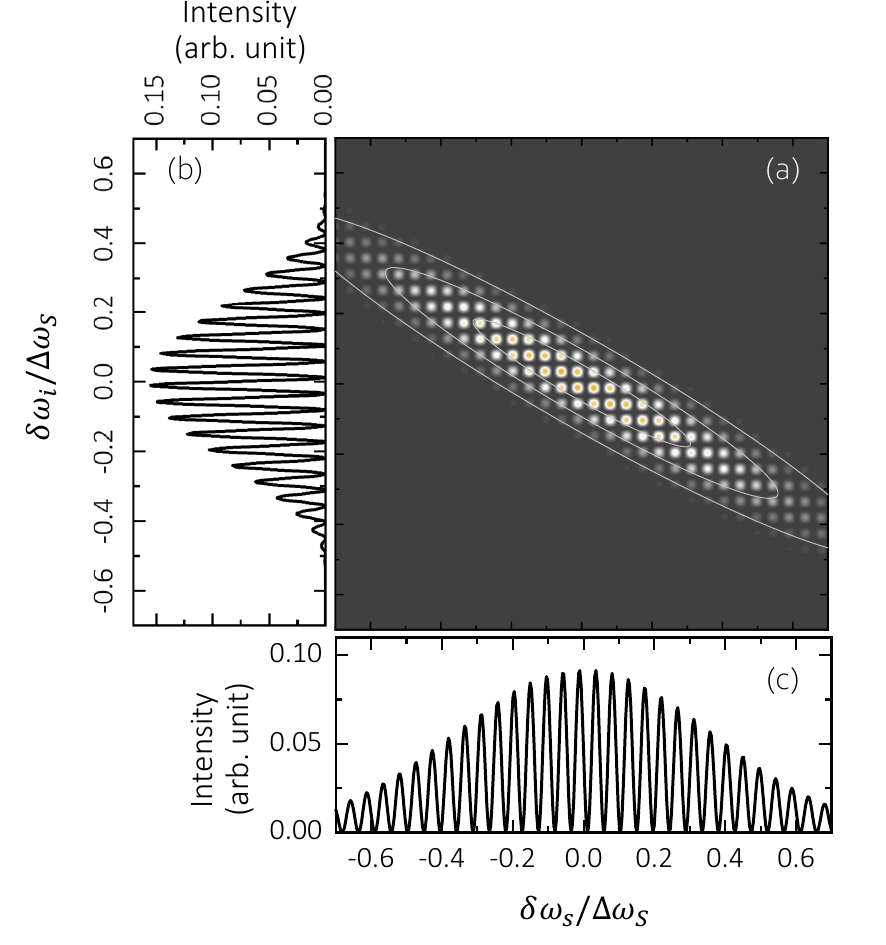}
\caption{Demonstration of modulation of SPDC JSI into a TF grid state. (a) Comparison of SPDC JSI (white curve) and TF grid state from Franson interference (color map). (b) Single-photon spectrum of $i'$-mode. (c) Single-photon spectrum of $s'$-mode. In the simulation, we use $f(\omega_s, \omega_i)=1/\sqrt{\mathcal{N}} e^{-2\ln(2)\omega_{-}^2/\Delta\omega_S^2/} e^{-2ln(2)\omega_{+}^2/\Delta\omega_p^2}$, where $\omega_{\pm}=\delta\omega_s \pm \delta\omega_i$,  $\delta\omega_{s} = (\omega_{s} - \omega_{s,0})/\alpha$, $\delta\omega_{i} = \omega_{i} - \omega_{i,0}$, $\omega_{s,i,0}$ is the central frequency of the signal and idler photons, $\alpha$ is the asymmetry parameter of JSA. $\Delta\omega_p$ represents the pump laser bandwidth, $\Delta\omega_S$ represents the SPDC single-photon bandwidth, and $\mathcal{N}$ is the normalized constant. The parameter set is: $\alpha$ = 1.7; $\Delta\omega_p/\Delta\omega_S=0.15$; $\tau_F=68/\Delta\omega_S$.}
\label{Fig_JSI_total}
\end{figure}

\begin{figure*}[ht]
\centering
\includegraphics[width=0.97\textwidth]{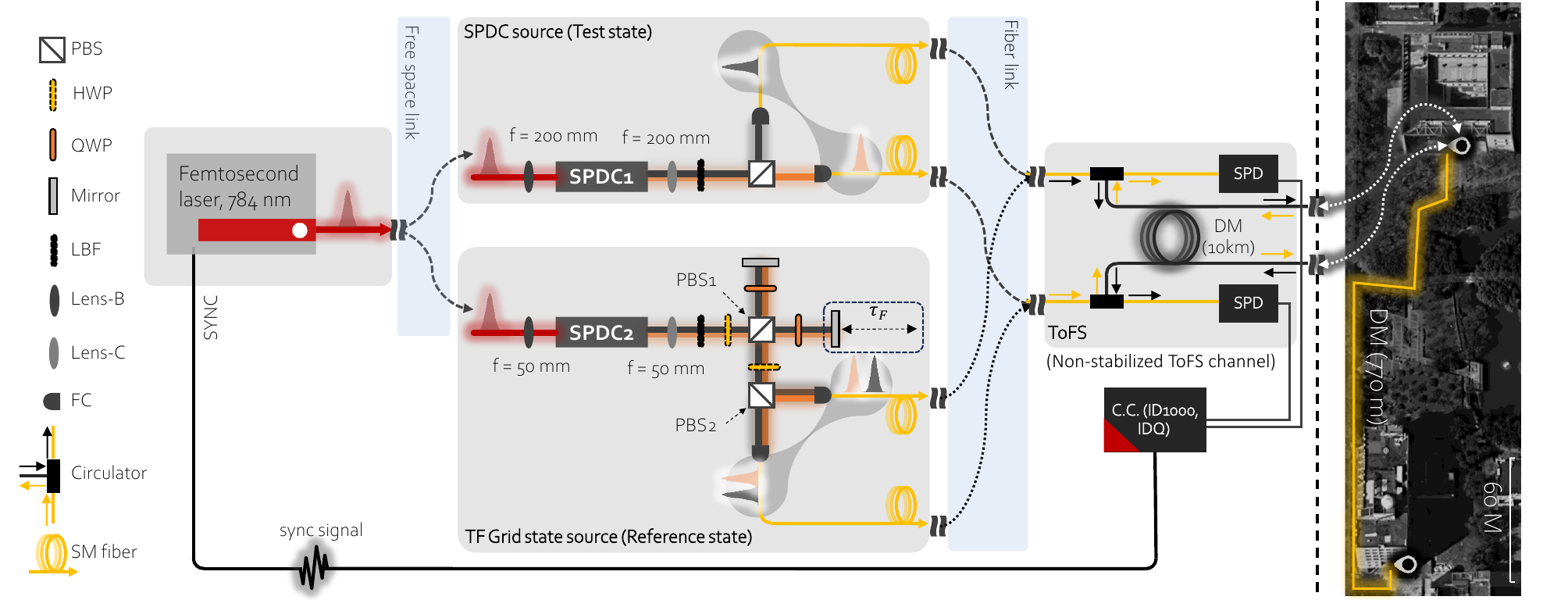}
\caption{Schematic diagram of the experimental setup. ToFS: Time-of-flight spectrometer. C.C.: Coincidence counting. DM: Dispersion medium.}
\label{setup_detail}
\end{figure*}

After the SPDC process, the generated photon pairs are sent into the polarization-based Franson interferometer \cite{park2018time,wang2025harnessing}. In the interferometer, a half-wave plate (HWP, set at $22.5^\circ$) first prepares the photon polarizations in the $\pm45^\circ$ basis. Subsequently, PBS1 equally splits the photon pairs into the long and short arms of the interferometer. Each arm consists of a quarter-wave plate (QWP, set at $45^\circ$) and a mirror, forming a double-pass configuration that ensures both photons exit from the same port of PBS1. The mirror in the long arm is mounted on a linear translation stage to control the Franson interference delay time, $\tau_F$. After exiting PBS1, the photon pairs pass through another HWP (set at $-22.5^\circ$), which again prepares the polarizations in the $\pm45^\circ$ basis. As a result, the photon pairs are equally split by PBS2 into two output ports and subsequently coupled into two single-photon detectors for coincidence counting measurement. 

According to the analysis in Ref. \cite{wang2025harnessing} and Supplementary Materials, the total two-photon JSA, $J_{A}^{\text{T}}$, output from the interferometer can be regarded as a superposition of the JSA of the two-photon entangled state (TPES) (corresponding to path-indistinguishable events), denoted as $J_{A}^{+}$, and the JSA of its satellite states, $J_{A}^{-}$ (corresponding to sideband events in the coincidence window), i.e. $J_{A}^{\text{T}}=J_{A}^{+}+J_{A}^{-}$. Those JSAs are given by
\begin{equation}
\begin{aligned}
J^{\pm}_A(\Omega_s,\Omega_i)=&\frac{1}{2}\left[f(\Omega_s,\Omega_i)e^{i\Omega_{s}\tau_H}\pm f(\Omega_i,\Omega_s)e^{i\Omega_{i}\tau_H}\right]\\
&\times\cos\left(\Omega_{\pm}\frac{\tau_F}{2}\right)e^{i\Omega_{+}(\tau_l+\tau_s)/2},
\label{JSA_pm}
\end{aligned}
\end{equation}
where $\Omega_{s,i}$ denotes the photon frequency on output modes $s'$ and $i'$ respectively, and $\Omega_{\pm}=\Omega_{s}\pm\Omega_{i}$. $\tau_{l,s}$ are the propagation time of long and short arms of Franson interferometer and $\tau_{F}=\tau_l-\tau_s$ is the time difference between long and short arms. 

Equation (\ref{JSA_pm}) shows that, unlike a conventional Franson interferometer \cite{jin2024spectrally}, we observe that the interferometer considered here incorporates the original JSA of SPDC and its frequency exchanged component with different phase that relate to the time delay of $\tau_H$. This structure, in fact, corresponds to HOM interference \cite{jin2016simple}. According to the relative phase in the superposition between $f(\Omega_s,\Omega_i)$ and $f(\Omega_i,\Omega_s)$, we find that $J_{A}^{+}$ corresponds to constructive interference, whereas $J_{A}^{-}$ corresponds to destructive interference. 

Furthermore, the JSA of both quantum states is influenced by Franson interference. For $J_{A}^{+}$, the interference occurs along the sum-frequency direction ($\Omega_{+}$), while for $J_{A}^{-}$, it occurs along the difference-frequency direction ($\Omega_{-}$). Therefore, by observing the Franson interference visibility of $J_{A}^{+}$, one can effectively quantify the frequency anti-correlation of the photon pairs, whereas $J_{A}^{-}$ provides a method to quantify the frequency correlation of the photon pairs via the Franson interference visibility.

It is worth noting that the modulation of $J_{A}^{\pm}$ by Franson interference along the $\Omega_{\pm}$ axes in the joint spectrum leads to an additional effect in the overall JSI, $J_{I}^{\text{T}} = \left| J_{A}^{+} + J_{A}^{-} \right|^{2}$. Specifically, the interference (i.e., cross terms) between $J_{A}^{\pm}$ gives rise to modulation along the $\Omega_{s,i}$ axes in the joint spectrum. As a result, $J_{I}^{\text{T}}$ forms a grid structure in the joint spectrum, which corresponds to the TF grid state.

However, we should not overlook the presence of HOM effects in this interferometer. If the SPDC operates under a degenerate condition, $f(\Omega_s,\Omega_i)$ and $f(\Omega_i,\Omega_s)$ will overlap in the joint spectrum. For $J_{A}^{+}$, constructive HOM interference is preserved, accompanied by Franson interference along the $\Omega_{+}$ direction. In contrast, the destructive HOM interference experienced by $J_{A}^{-}$ suppresses its contribution in the degenerate region. As a result, $J_{I}^{\text{T}}$ in the degenerate region is dominated by $J_{A}^{+}$, which further leads to an incomplete TF grid structure. 

To suppress these effects while preserving the modulation of the $J_{I}^{\text{T}}$ induced by Franson interference, we consider a strongly non-degenerate condition, which makes the HOM interference between $f(\Omega_s,\Omega_i)$ and $f(\Omega_i,\Omega_s)$ negligible (see Supplementary Materials for the detailed discussion on the degenerate case ). 

In this regime, the two contributions evolve independently under Franson interference, without significant cross-interference, and jointly dominate $J_{I}^{\text{T}}$. In the joint spectrum, these two contributions are symmetrically distributed along the $\Omega_s=\Omega_i$ axis. In practice, we selectively analyze one of these contributions for subsequent distortion detection and correction. Based on the above conditions, $f(\Omega_i,\Omega_s)$ can be neglected, and the resulting output JSI $J^{\text{T}}_{I}(\Omega_s,\Omega_i)$ can be expressed as
\begin{equation}
\begin{aligned}
J^{\text{T}}_I(\Omega_s,\Omega_i)&=\left|f(\Omega_s,\Omega_i)\right|^2\cos^2\left(\frac{\Omega_s\tau_F}{2}\right)\cos^2\left(\frac{\Omega_i\tau_F}{2}\right).
\label{JSI_total}
\end{aligned}
\end{equation}

Equation (\ref{JSI_total}) shows that the JSI of the SPDC source, $|f(\Omega_s,\Omega_i)|^2$, is modulated by a cosine-squared function along the $\Omega_s$ and $\Omega_i$ directions with a fixed period of $1/\tau_F$. As a result, the JSI $|f(\Omega_s,\Omega_i)|^2$ is no longer a continuous distribution but is mapped onto a discrete lattice structure. This grid structure serves as an intrinsic coordinate system in the frequency domain.

In Fig. \ref{Fig_JSI_total}, we present the JSI obtained from the considered interferometer under a strongly non-degenerate condition. In this illustration, one can clearly observe the periodic modulation of $|f(\Omega_s,\Omega_i)|^2$ (the white curve in Fig. \ref{Fig_JSI_total}(a)) induced by Franson interference, forming a TF grid state. 

Furthermore, as shown in Fig. \ref{Fig_JSI_total} (a), (b), and (c), these periodic structures provide well-defined reference points in the frequency domain. If the TF grid state propagates through a channel or measurement system that introduces coordinate deformation, the reconstructed TF distribution becomes distorted. Under such conditions, the TF grid state can serve as a built-in reference for identifying the deformation and enabling corresponding post-processing correction. This resource thus holds strong potential for applications in frequency-domain quantum information processing \cite{fabre2020generation,jin2024spectrally,fabre2023teleportation}.

\section{Experimental setup}\label{secIII}

Based on the theoretical analysis above, we design an experiment to demonstrate the generation and application of TF grid states for distortion detection in optical fiber networks. The experiment consists of two independent laboratories (the Photon Source Lab and the Node Lab), which are connected by two deployed campus optical fibers with a length of approximately 770 m, as shown in Fig. \ref{setup_detail}. In the Node Laboratory, two deployed campus optical fibers are connected to form a loop configuration, allowing the photons sent from the Photon Source Laboratory to propagate through an outdoor field environment and be redirected back to the Photon Source Laboratory, thereby simulating a realistic fiber-based transmission scenario.

Two photon sources are prepared in the Photon Source Laboratory (Fig. \ref{setup_detail}). First, a photon-pair source based on a nonlinear crystal (SPDC1, type-II MgO-doped congruent lithium niobate, 25 mm, $125^\circ\mathrm{C}$, non-degeneracy, HCP) is constructed to generate frequency-anticorrelated photon pairs, which serves as the \textit{test state} for simulating the transmission of frequency-entangled photon pairs in a non-ideal fiber-based measurement channel. Second, another photon-pair source (SPDC2, type-II MgO-doped congruent lithium niobate, 1 mm, $145^\circ\mathrm{C}$, non-degeneracy, HCP) is sent through a polarization-based Franson interferometer to generate a TF grid state as a \textit{reference state}, which serves as a reference to probe distortions in a non-ideal fiber-based measurement channel.

Both sources are pumped by a pulsed laser (FemtoFiber smart 780, TOPTICA) with a repetition rate of 80 MHz, a pulse duration of 81 fs, a central wavelength of 784 nm, a spectral bandwidth of approximately 5 nm, and an average power of 60 mW. 

For SPDC1, the pump beam is coupled through free space and focused into the crystal by a lens ($f = 200$ mm). The generated photon pairs (with signal and idler wavelengths of 1620 nm and 1520 nm, respectively) are collimated by a lens ($f = 200$ mm), separated by a PBS, and coupled into optical fibers, thereby completing the preparation of the test state.

For SPDC2, the generated photon pairs are sent into a polarization-based Franson interferometer (with arm length difference around 0.437 mm) as described in Sec. \ref{secII} to establish the TF grid state and complete the preparation of the reference state.

After the quantum states are prepared, the photon pairs of the test state and the reference state are individually connected via optical fibers to a pair of circulators. A 10 km indoor single-mode dispersive fiber and a 770 m deployed outdoor loop fiber are connected between the two circulators, forming a fiber transmission link. Finally, the input photon pairs counter-propagate through the fiber link and exit from the opposite circulator.

In this configuration, the 10 km and 770 m fibers serve two roles. First, they can be regarded as a dispersive medium, where different photon frequencies are mapped to different photon arrival times. Therefore, this setup can be employed as a ToFS to reconstruct the JSI of the photon pairs \cite{wang2025harnessing,davis2017pulsed}. To reconstruct the JSI of both the test state and the reference state, the output photons from two circulators are detected by two single-photon detectors (ID230, ID Quantique) and the signals are sent to a time controller (ID1000, ID Quantique) to record the time tags, $\{t^m_s,t^m_i\}$. Furthermore, the synchronization signal of the pulsed laser is sent to a time controller as a clock. By analyzing the three-fold coincidences among the signal photon, idler photon, and laser synchronization signal, the JSIs of the quantum states are reconstructed. 

It is important to note that, through the ToFS process, photon frequencies are mapped onto arrival times ($\Omega_{s,i}\rightarrow t^m_{s,i}$). Therefore, the temporal axes of the measured JSI provide an experimentally accessible representation of the photon-frequency coordinates. 

Secondly, it should be emphasized that the fiber transmission channel is not actively stabilized and contains a deployed outdoor fiber segment operating in a realistic environment. As a result, the frequency-to-time mapping established by the ToFS is subject to distortions arising from environmental fluctuations, non-ideal dispersive responses, and other channel-dependent effects. These distortions introduce coordinate deformation into the measured temporal distribution, which is subsequently transferred to the reconstructed JSI. Therefore, the fiber link serves not only as the dispersive medium required for ToFS, but also as a non-ideal measurement channel that distorts the reconstructed TF distributions.

Based on the discussions above, we have established a non-ideal measurement environment, together with an independent frequency-entangled test state and a TF grid state serving as a reference probe. The key objective of this work is to investigate whether the intrinsic periodic structure of the TF grid state can be exploited to reconstruct channel-induced coordinate distortions without prior knowledge of the underlying perturbation mechanism. More importantly, we examine whether the reconstructed distortion model can be generalized beyond the reference state itself and applied to recover an independent TF quantum state propagating through the same transmission channel. The following sections present the experimental results and demonstrate the reconstruction and correction procedures developed in this work.


\begin{figure*}[ht]
\centering
\includegraphics[width=0.9\textwidth]{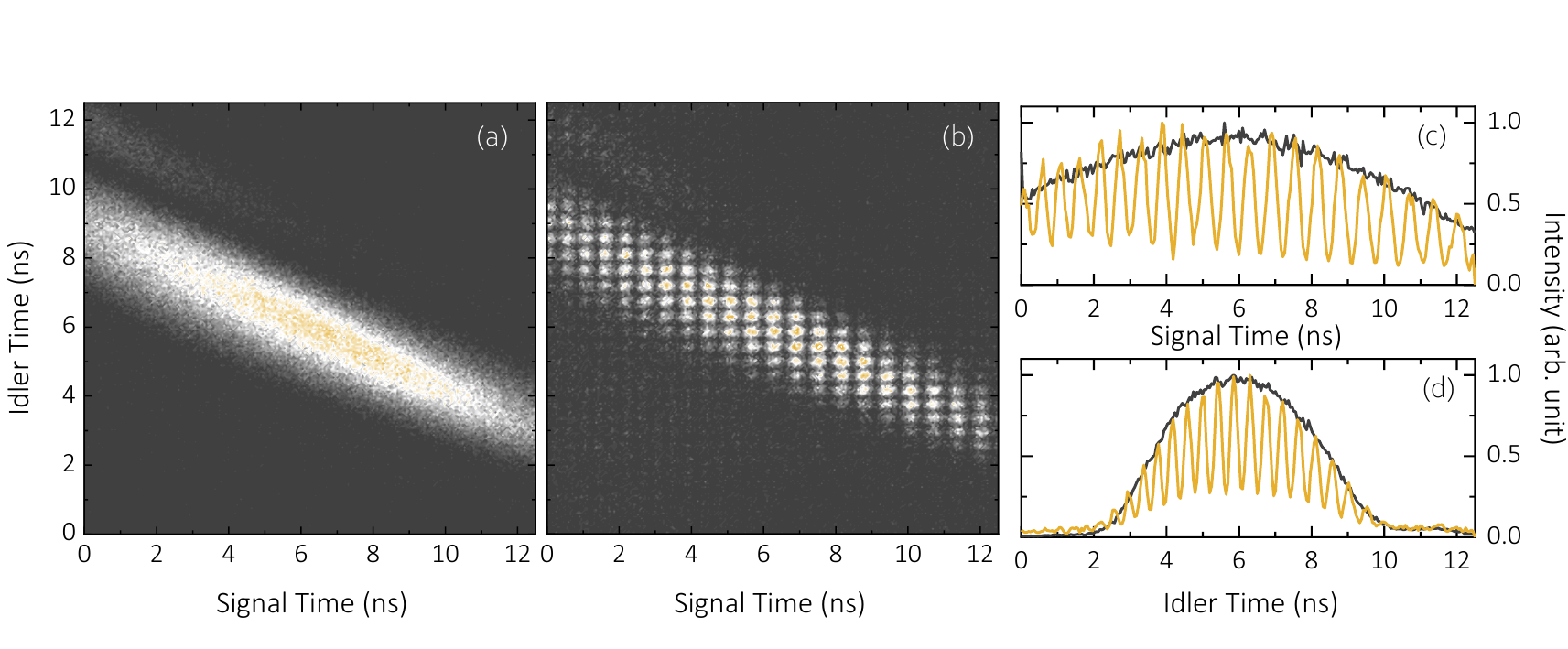}
\caption{Demonstration of the JSI and marginal spectra of the TF grid state. 
(a) and (b) are the JSI of SPDC2 w/o and w/ modulation by Franson interference. (c) and (d) show the marginal spectra of SPDC2 w/o (black) and w/ (yellow) modulation by Franson interference.}
\label{Grid_data}
\end{figure*}

\section{Results and Discussion}\label{secIV}

Following the workflow illustrated in Fig. \ref{Flow}, the prepared test state and reference state were transmitted through the fiber system to demonstrate the measurement of distorted TF distributions and their subsequent correction under a non-ideal fiber-based measurement channel.

\subsection{JSI measurement of TF grid state}\label{secIV A}

Based on the experimental setup, our first step is to generate a TF grid state using the polarization-based Franson interferometer and employ it as a reference state for characterizing the channel-induced distortion. 

Before generating the TF grid state, the original JSI of SPDC2 was first characterized using the ToFS system without passing through the interferometer, as shown in Fig. \ref{Grid_data}(a). It can be observed that SPDC2 exhibits a broad two-photon bandwidth due to its shorter crystal length. This broader spectral bandwidth is beneficial for the subsequent interferometric modulation over the joint spectrum, allowing a larger number of resolvable grid points within the accessible spectral range.

Finally, after the photon pairs generated from SPDC2 were modulated by the polarization-based Franson interferometer and reconstructed through the ToFS system, a clearly resolved TF grid state JSI was observed, as shown in Fig. \ref{Grid_data}(b).

By comparing Fig. \ref{Grid_data}(a) and Fig. \ref{Grid_data}(b), the modulation effect introduced by the interferometer on the original SPDC2 JSI can be clearly observed. The interferometer effectively generates orthogonal modulations along both the signal and idler axes, transforming the continuous distribution into a discrete grid-like structure, showing good agreement with the theoretical model.

Furthermore, Fig. \ref{Grid_data}(c) and Fig. \ref{Grid_data}(d) present the corresponding projections along each frequency axis, where the discrete spectral features are clearly observed while still following the original spectral envelope.

Although the grid structures presented in Fig. \ref{Grid_data}(b)-(d) do not exhibit obvious distortions visually, the following analysis will demonstrate that these grid structures are capable of sensitively resolving such distortions and further reconstructing the distortion behavior introduced by the channel. 

\subsection{Reconstruction of the correction mapping from grid-point displacements}\label{secIV B}

\begin{figure*}[ht]
\centering
\includegraphics[width=0.89\textwidth]{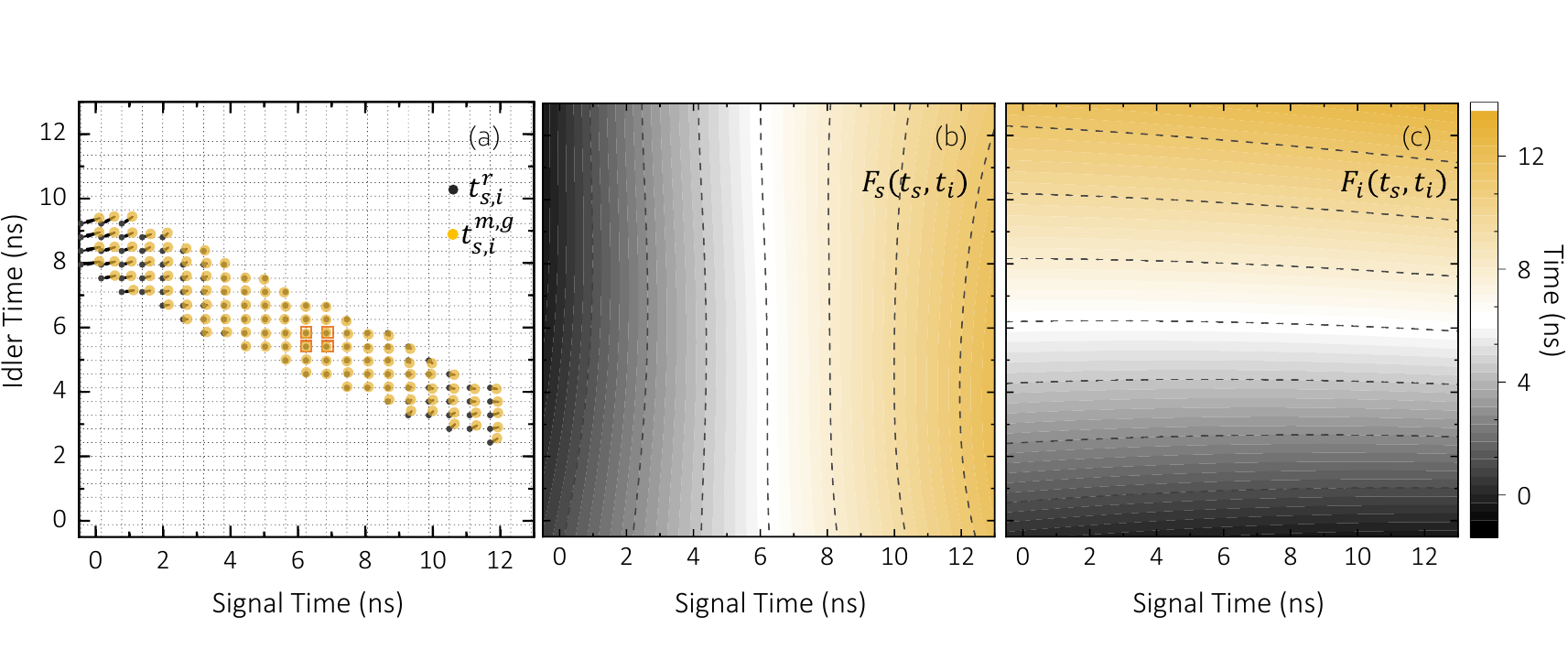}
\caption{Reconstruction of the correction mapping using the GPR model. (a) Comparison between the measured grid points and the corresponding ideal grid points. The yellow circles represent the experimentally measured grid positions, $\{t_s^{m,g},t_i^{m,g}\}$. The four grid points enclosed by the red square are used as the reference points for constructing the ideal periodic grid system, which is indicated by the black dashed lines. The black dots represent the corresponding ideal coordinates $\{t_s^r,t_i^r\}$ of each measured grid point, while the displacement between the measured and ideal positions is indicated by the black arrows. (b) and (c) Correction mapping functions obtained from the trained GPR model for the signal and idler channels ($F_s(t_s,t_i)$ and $F_i(t_s,t_i)$), respectively. The horizontal and vertical axes correspond to the experimentally measured signal and idler arrival times, while the color map represents the output value of the corresponding mapping function at each coordinate.}
\label{ML_correctiona}
\end{figure*}

Through the measurements described above, sufficient information is now available to investigate the coordinate deformation introduced by the channel into the measured TF distribution. To analyze these effects, we first localize the measured grid points. To do so, we perform localized segmentation on each grid point of the measured reference state. The coincidence counts of the photon pairs within each segmented region (defined as the region containing each wave packet) are then used as weighting factors to calculate the averaged position of each grid point. As a result, a set of measured grid-point coordinates, $\{t_s^{m,g},t_i^{m,g}\}$, is obtained, whose averaged distribution is represented by the yellow circles in Fig. \ref{ML_correctiona}(a).

Next, in order to determine whether the measured grid points exhibit displacement, it is necessary to establish a theoretically expected grid system on the JSI. Here, four grid points located near the center region of the overall grid distribution were first selected as the basis for constructing the ideal grid structure, denoted as a set of $\{t_{s,0}^{m,g},t_{i,0}^{m,g}\}$ and indicated by the red square in Fig. \ref{ML_correctiona}(a). Using these four measured grid-point coordinates, a rectangular structure was fitted and subsequently extended according to its geometrical periodicity to generate an ideal periodic grid system, which serves as the ideal coordinate reference of the TF grid state, as illustrated by the black dashed lines in Fig. \ref{ML_correctiona}(a). Based on this constructed grid system, the theoretically expected coordinates corresponding to each measured grid point can then be identified by progressively extending from the center toward the outer regions. The resulting reference coordinates set, $\{t_s^r,t_i^r\}$, are represented by the black points in Fig. \ref{ML_correctiona}(a).

As shown in Fig. \ref{ML_correctiona}(a), the measured TF grid-state points gradually deviate from the ideal coordinates $\{t_s^r,t_i^r\}$ as their distance from $\{t_{s,0}^{m,g},t_{i,0}^{m,g}\}$ increases, as indicated by the black arrows. Importantly, these systematic displacements suggest that the channel modifies the coordinate correspondence between the ideal and measured TF distributions. In other words, the observed distortion can be interpreted as a channel-induced coordinate deformation that maps reference coordinates to measured coordinates. This result further demonstrates the capability of the TF grid state to detect channel-induced mapping distortions.

The correction task is to learn the inverse relation, namely a mapping from measured coordinates back to reference coordinates. If the corresponding correction mapping can be reconstructed, it becomes possible to recover experimentally measured TF distributions toward their ideal undistorted forms.

In this work, we do not assume any specific physical mechanism responsible for the observed distortion at the initial stage (the possible physical origins will be discussed later). Instead, the entire transmission channel is treated as a black-box system in order to simulate the influence and correction of arbitrary unknown distortions on TF quantum states. To reconstruct the correction mapping between the measured coordinates $\{t_s^{m,g},t_i^{m,g}\}$ and the ideal grid coordinates $\{t_s^r,t_i^r\}$, we introduce machine learning as a data-driven reconstruction approach. In this problem, $\{t_s^{m,g},t_i^{m,g}\}$ are treated as the input training data, while $\{t_s^r,t_i^r\}$ serve as the corresponding target outputs. Here, Gaussian process regression (GPR) is chosen \cite{williams2006gaussian} to construct the model since it provides a smooth nonlinear mapping while maintaining robustness under sparse training data. 

By training the GPR model, we learn the correction mapping between the measured and ideal coordinates, thereby establishing the correction mapping between arbitrary input temporal coordinates $\{t_s,t_i\}$ and their corresponding outputs $\{F_s(t_s,t_i),F_i(t_s,t_i)\}$, where $F_{s,i}(t_s,t_i)$ represents the output of signal mode and idler mode, respectively. The resulting model provides a correction mapping that compensates for the nonlinear coordinate deformation induced by the transmission channel and serves as the basis for the subsequent JSI correction and reconstruction. The corresponding results of the trained GPR model for $F_s(t_s,t_i)$ and $F_i(t_s,t_i)$ under the corresponding signal- and idler-time coordinates are shown in Fig. \ref{ML_correctiona}(b) and (c). 

Figure \ref{ML_correctiona}(b) and (c) further reveal the correction mapping between the measured and corrected temporal coordinates. Although both $F_s(t_s,t_i)$ and $F_i(t_s,t_i)$ exhibit a dominant positive correlation along their respective signal and idler axes, reflecting the fundamental operating principle of the ToFS system, a residual dependence on the cross coordinate can still be observed. The resulting deformation therefore cannot be fully decomposed into two independent one-dimensional mappings, suggesting that the distortion introduced by the channel should be treated as a two-dimensional coordinate deformation process.

Interestingly, these results can be interpreted as an effective two-dimensional coordinate deformation field acting on the measured TF distribution during transmission through the channel. Under this interpretation, the experimentally measured temporal coordinates are distorted by the field, leading to the observed distortion of the reconstructed TF distribution. Therefore, by reconstructing this deformation field, it becomes possible to directly visualize the influence of the transmission channel on arbitrary TF quantum states.

\begin{figure}[t]
\centering
\includegraphics[width=0.37\textwidth]{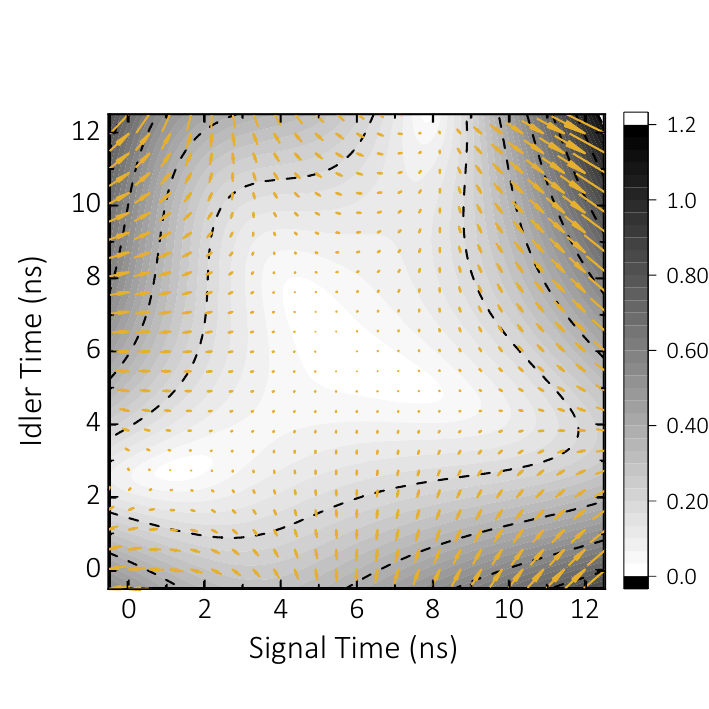}
\caption{Reconstructed effective channel-induced deformation field. The arrows represent the local displacement vectors  $\{t_s-F_s(t_s,t_i);t_i-F_i(t_s,t_i)\}$, while the color map denotes the displacement magnitude. The increasing displacement away from the central reference region indicates a spatially varying nonlinear coordinate deformation rather than a global timing offset.
}
\label{Vector_field}
\end{figure}

To visualize the deformation field more directly, we convert the learned correction mapping into an effective two-dimensional displacement field from the results shown in Fig. \ref{ML_correctiona}(b) and (c), by treating $\{t_s-F_s(t_s,t_i);t_i-F_i(t_s,t_i)\}$ as the local displacement vector at each point in the two-dimensional temporal coordinate space. Using this approach, the reconstructed deformation field can be visualized as shown in Fig. \ref{Vector_field}, where the arrows indicate the local displacement direction associated with the coordinate transformation of a TF quantum state propagating through the transmission channel, while the color map represents the corresponding displacement magnitude.

It can be observed that the distortion introduced by the channel is not uniformly distributed over the coordinate space. The displacement magnitude gradually increases as the distance from the center of the reference grid ($\{t_{s,0}^{m,g},t_{i,0}^{m,g}\}$) increases. This result indicates that the channel-induced mapping distortion cannot be described by a simple global timing offset, but instead varies across the coordinate space. Furthermore, the observed behavior suggests that the frequency-to-time mapping becomes increasingly nonlinear away from the central region of the measurement space.


It should be noted that the observed non-separable mapping feature does not necessarily imply the existence of additional physical interactions between the signal and idler channels. Such behavior may also arise from the finite sampling range of the experimentally accessible grid-state distribution used for training the GPR model. Furthermore, in the present experiment, this nonlinear coordinate deformation is most likely associated with the nonideal frequency-to-time mapping of the dispersive fiber link, including higher-order dispersion contributions. 

Nevertheless, the reconstruction procedure itself does not rely on this physical assumption, because the mapping is learned directly from the measured TF grid state. The reconstructed deformation field therefore provides a direct visualization of the channel-induced mapping distortion and establishes the basis for the subsequent correction procedure.

\begin{figure*}[ht]
\centering
\includegraphics[width=0.89\textwidth]{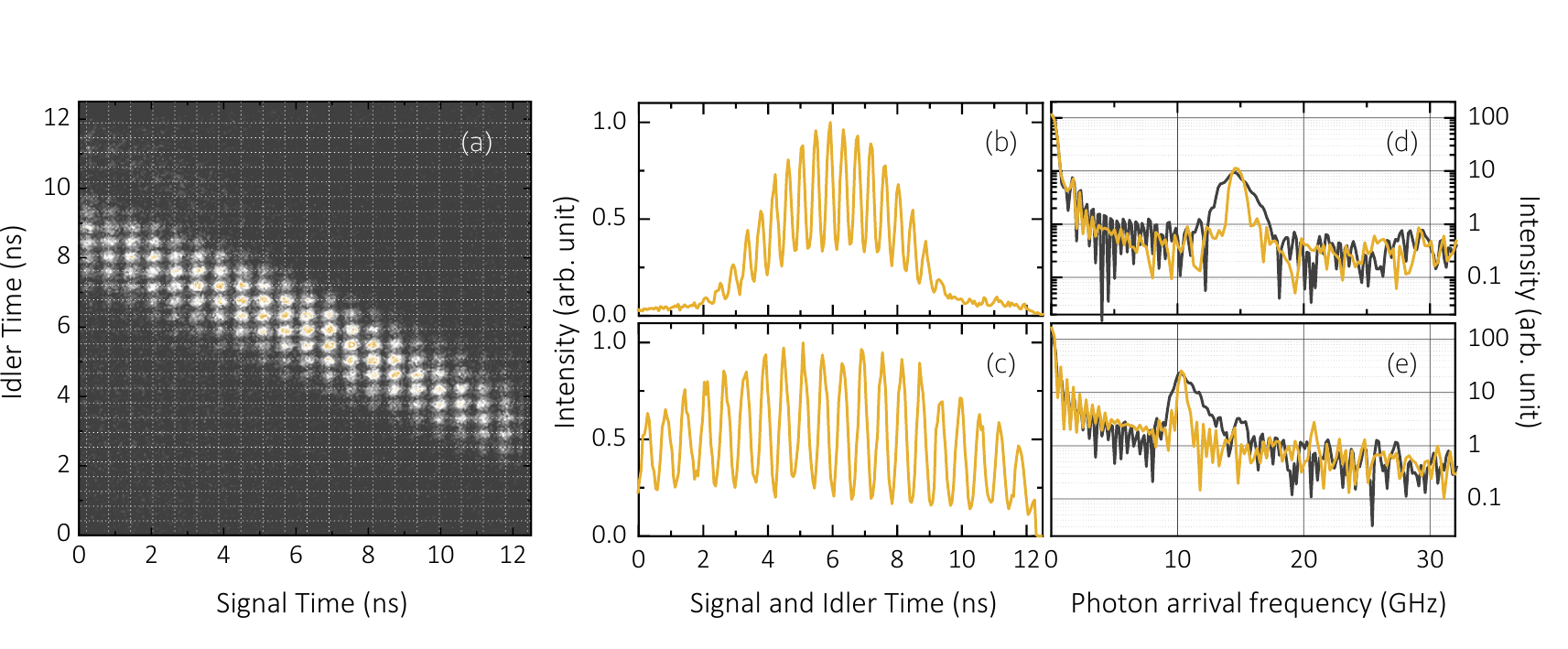}
\caption{Reconstruction and analysis of the TF grid state. 
(a) Reconstructed JSI of the TF grid state, where the white grid lines denote the reference grid system. (b) and (c) show the marginal spectra along the signal and idler axes, respectively. (d) and (e) show the Fourier spectra of the marginal distributions in (b) and (c), respectively. Yellow and black curves correspond to the corrected and uncorrected results.}
\label{Grid correction}
\end{figure*}

\subsection{TF grid-state reconstruction and correction}\label{secIV C}

Using the correction mapping, it becomes possible to correct the measured TF distributions and recover their corresponding ideal distributions.

Since the ideal periodic structure of the TF grid state is known a priori, it provides a convenient benchmark for validating the reconstructed mapping before applying it to arbitrary TF quantum states. Therefore, we first apply the correction mapping ($F_s(t_s^m,t_i^m)$ and $F_i(t_s^m,t_i^m)$) to the measured TF grid state. The corrected result is shown in Fig. \ref{Grid correction}(a), where the white grid lines represent the reference grid system. It can be observed that, after correction, the reconstructed grid points are well aligned with the reference grid coordinates, indicating that the distortion in the measured TF distribution has been effectively compensated.

To further verify the validity of the correction mapping, the corrected TF grid state shown in Fig. \ref{Grid correction}(a) is projected onto the signal and idler axes by integrating over the corresponding orthogonal coordinate. The resulting marginal spectra are presented in Fig. \ref{Grid correction}(b) and (c), respectively.

Furthermore, Fourier analysis is performed on the corresponding marginal spectra to evaluate the periodicity of the reconstructed grid structure. The resulting Fourier spectra are shown as the yellow curves in Fig. \ref{Grid correction}(d) and (e), while the black curves correspond to the Fourier spectra obtained before correction.

In the uncorrected case, the distortion makes the grid distribution less periodic. After applying the correction mapping ($F_s(t_s^m,t_i^m)$ and $F_i(t_s^m,t_i^m)$), the displaced grid points are realigned toward their ideal positions, restoring the periodic structure of the TF grid state. As a result, the corrected spectrum becomes more concentrated, resulting in a narrower Fourier spectrum than the uncorrected cases. This analysis provides a more intuitive verification of the effectiveness of the reconstructed coordinate mapping and demonstrates its capability to recover the underlying periodic structure of the TF grid state.

\begin{figure}[t]
\centering
\includegraphics[width=0.37\textwidth]{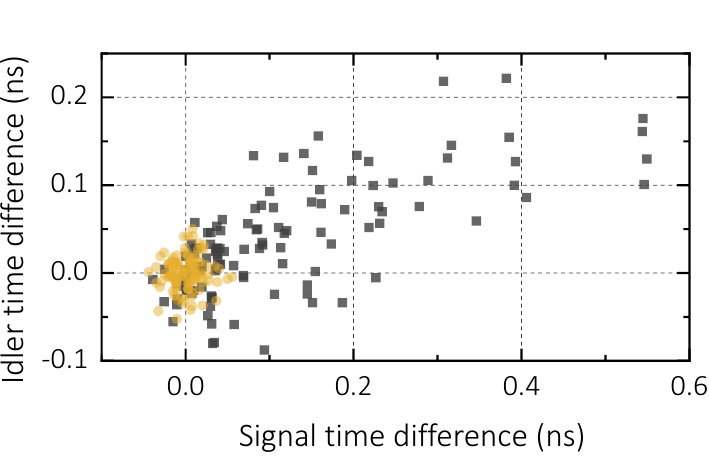}
\caption{Distribution of grid-point displacement before and after correction. The gray squares and yellow dots represent the uncorrected and corrected grid-point displacements relative to their corresponding reference coordinates, respectively.}
\label{grid_error}
\end{figure}

On the other hand, to quantitatively evaluate the performance of the correction mapping, we further compare the displacement distributions before and after correction, as shown in Fig. \ref{grid_error}. For each grid point, the displacement is defined relative to its corresponding reference coordinate $\{t_s^r,t_i^r\}$. The uncorrected and corrected displacement vectors are therefore given by $\{t_s^{m,g}-t_s^r,t_i^{m,g}-t_i^r\}$ and $\{F_s(t_s^{m,g},t_i^{m,g})-t_s^r,F_i(t_s^{m,g},t_i^{m,g})-t_i^r\}$, respectively. It can be clearly observed that the corrected grid points exhibit a significantly more concentrated distribution around the origin. 

To quantitatively characterize this improvement, we calculate the radial displacement of each grid point and evaluate the standard deviation of its distribution. The standard deviation is reduced from 0.134 ns before correction to 0.012 ns after correction, corresponding to an approximately 11-fold reduction in the radial displacement standard deviation. This result quantitatively confirms that the reconstructed correction mapping accurately compensates for the channel-induced coordinate deformation and provides an effective basis for recovering distorted TF distributions.

\begin{figure*}[t]
\centering
\includegraphics[width=0.89\textwidth]{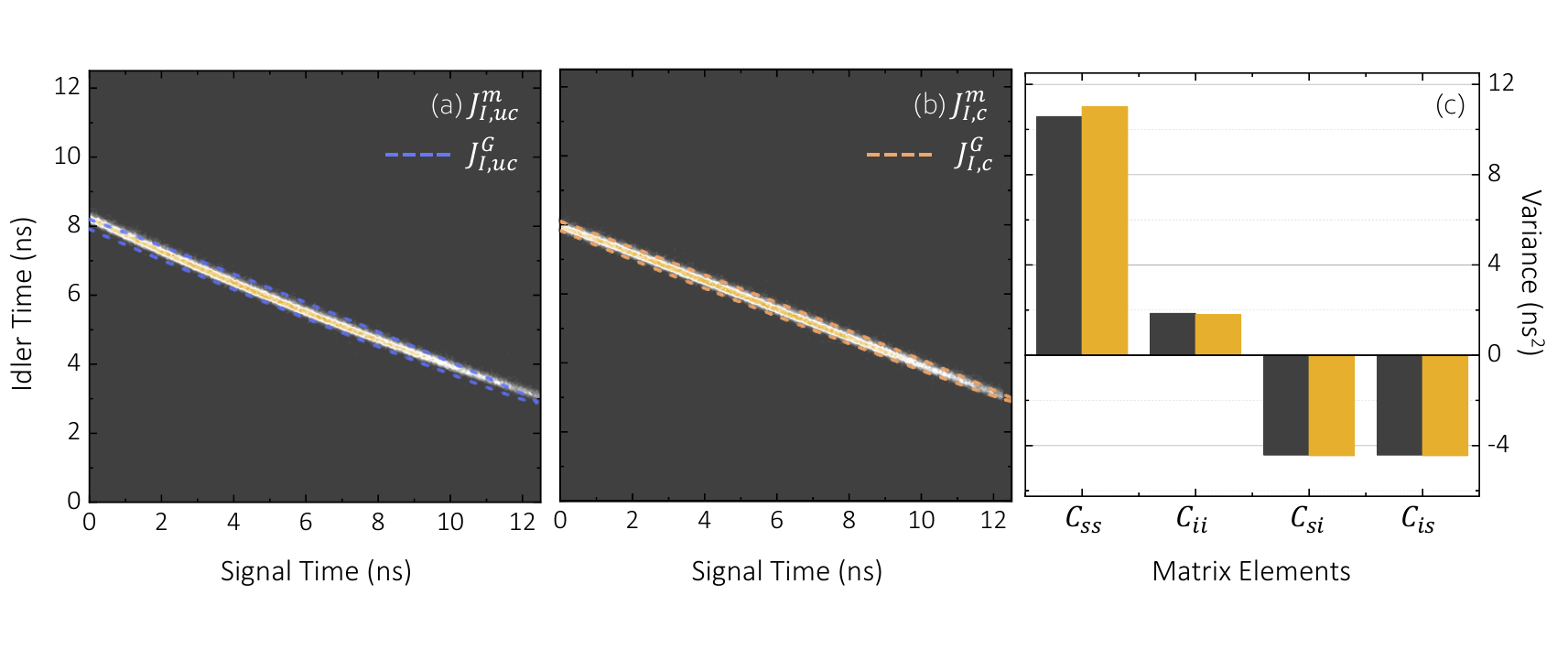}
\caption{Analysis of measured and corrected JSIs of the test state. (a)  Uncorrected measured JSI $J_{I,uc}^m$. (b) Corrected JSI, $J_{I,c}^m$ obtained using the reconstructed coordinate mapping. The blue and orange contours represent the corresponding Gaussian-approximations JSIs, $J_{I,uc}^G$ and $J_{I,c}^G$, respectively. (c) Covariance-matrix elements extracted from the uncorrected and corrected JSIs. Gray and yellow bars denote the uncorrected and corrected cases, respectively.
}
\label{SPDC correction}
\end{figure*}

\subsection{Independent frequency-entangled test-state reconstruction and correction}\label{secIV D}

After validating the reconstructed correction mapping using the TF grid state, here, we demonstrate the correction of an independent frequency-entangled state using the correction mapping reconstructed from the TF grid state. 

Before presenting the correction results, the test state was first transmitted through the fiber link, and its JSI was reconstructed using the ToFS system. The uncorrected measured JSI, $J_{I,uc}^m$, is shown in Fig. \ref{SPDC correction}(a). A clear frequency anticorrelation can still be observed, indicating that the underlying SPDC spectral correlation is roughly preserved after long-distance transmission.

However, upon closer inspection, $J_{I,uc}^m$ exhibits noticeable distortions in its distribution. These deviations are consistent with perturbations in the frequency-to-time mapping of the ToFS system induced by the transmission channel. As a result, the reconstructed TF distribution becomes distorted, leading to the deformation of the reconstructed JSI.

Unlike the TF grid state used to establish the mapping model, $J_{I,uc}^m$ possesses a continuous frequency-anticorrelated JSI distribution and therefore serves as an independent test state for evaluating the general applicability of the proposed framework. We then apply the same correction mapping to $J_{I,uc}^m$ and further the corrected JSI, the result denoted by
$J_{I,c}^{m}$, is shown in Fig. \ref{SPDC correction} (b). One can clearly observe that the distortion present in $J_{I,uc}^{m}$ is substantially reduced after applying the correction mapping.

To quantitatively evaluate the distortion introduced by the transmission channel and the performance of the correction method, it is first necessary to establish a reference model for the JSI that is expected to be observed experimentally for the frequency-entangled test state. In theory, the JSI of a frequency-entangled state generated from a bulk SPDC crystal is primarily determined by the phase-matching function of the crystal and the pump spectrum, resulting in a characteristic sinc-like distribution \cite{gayer2008temperature}.

However, in a practical ToFS, the finite timing resolution of the single-photon detectors and the time-tagging electronics introduces additional temporal uncertainty (150.11 ps). The combined timing jitter smooths the fine spectral features of the ideal JSI and broadens the measured distribution. As discussed in detail in the Supplementary Materials, numerical simulations show that the detector-jitter-broadened JSI can be accurately approximated by a Gaussian model, with a fidelity exceeding 95.6\% over the experimental parameter range considered in this work. Therefore, under realistic experimental conditions, the measured JSI is theoretically expected to exhibit a Gaussian-like profile and can be well approximated by a Gaussian model.

Based on this observation, Gaussian approximations are constructed from the covariance matrices of both the uncorrected and corrected JSIs and used as experimentally motivated reference models. By comparing each measured JSI with its corresponding Gaussian approximation, the deviation of the measured TF distribution from the expected Gaussian-like correlation geometry can be quantified, thereby providing a metric for evaluating the effectiveness of the correction procedure. 

To construct the corresponding Gaussian approximation in both cases of uncorrected and corrected, we evaluate the covariance matrix (CM) for both $J_{I,uc}^{m}$ and $J_{I,c}^{m}$. The extracted mean values and covariance matrix are then used to construct the corresponding Gaussian approximation, denoted by $J_{I,(u)c}^{G}$. For a given CM, the corresponding Gaussian distribution can be written as
\begin{equation}
J_{I,(u)c}^{G}(\mathbf{t})=\frac{\exp\left[
-\frac{1}{2}(\mathbf{t}-\boldsymbol{\mu}_{(u)c})^{T}\mathbf{C}_{(u)c}^{-1}(\mathbf{t}-\boldsymbol{\mu}_{(u)c})\right]}{2\pi \sqrt{|\mathbf{C}_{(u)c}|}},
\label{Gaussian_JSI}
\end{equation}
where
\begin{equation}
\mathbf{t}=
\begin{pmatrix}
t_s\\
t_i
\end{pmatrix},
\boldsymbol{\mu}_{(u)c}=
\begin{pmatrix}
\mu_{s,(u)c}\\
\mu_{i,(u)c}
\end{pmatrix},
\mathbf{C}_{(u)c}=
\begin{pmatrix}
C_{ss}^{(u)c} & C_{is}^{(u)c}\\
C_{si}^{(u)c} & C_{ii}^{(u)c}
\end{pmatrix}
\end{equation}
represent the time coordinate vector, mean vectors, and covariance matrices of the measured JSIs, respectively. In Fig. \ref{SPDC correction}(c), the analyzed CMs before and after correction are shown as the gray and yellow bars, respectively. The corresponding correlation coefficients and aspect ratios remain comparable before and after correction, indicating that the correction procedure preserves the intrinsic frequency-anticorrelation structure of the state. Furthermore, based on these CMs and Eq. \ref{Gaussian_JSI}, the estimated Gaussian JSIs for the uncorrected and corrected cases are plotted as the blue and orange dashed contours in Fig. \ref{SPDC correction}(a) and (b), respectively. 

Based on the above results, we can now quantify the deviation of the measured TF distribution from its expected Gaussian-like correlation geometry before and after correction. To this end, we define the Gaussian-shape fidelity between the measured JSI and its corresponding Gaussian approximation as
\begin{equation}
F^{G}_{(u)c}(J_{I,(u)c}^{G},J_{I,(u)c}^{m}) =\left|\int\int \sqrt{J_{I,(u)c}^{G}J_{I,(u)c}^{m}}\,dt_sdt_i\right|^2.
\end{equation}
A higher value of $F^{G}_{(u)c}$ indicates that the measured TF distribution is more consistent with the expected Gaussian-like correlation geometry. After the estimation, the Gaussian-shape fidelity increases significantly from 76.2\% before correction to 90.0\% after correction. This substantial improvement demonstrates that the correction mapping successfully captures the dominant channel-induced coordinate deformation. More importantly, the correction is achieved using a correction mapping learned exclusively from the TF grid state and subsequently applied to an independent frequency-entangled state without additional optimization, highlighting the transferability of the proposed framework.

\section{Summary and Conclusion}\label{secV}

In this work, we experimentally demonstrated the use of a TF grid state as a reference resource for probing and correcting TF-distribution distortions caused by channel-induced coordinate deformation. To the best of our knowledge, this represents the first experimental realization of TF-grid-state-assisted correction of TF distributions in a realistic measurement environment.

By exploiting the intrinsic periodic grid structure of the TF grid state in the joint spectrum, the displacement of individual grid points can identify the nonlinear coordinate deformation introduced by an imperfect transmission channel and reconstruct a corresponding correction mapping. These results demonstrate that TF grid states can provide an intrinsic frequency-domain coordinate reference for characterizing distortions in realistic measurement environments.

Using the measured TF grid state, a correction mapping was reconstructed without assuming any prior knowledge of the underlying perturbation mechanism. The reconstructed correction mapping was first validated through the recovery of the TF grid state itself, yielding an approximately 11-fold reduction in the residual coordinate deviation. More importantly, the same mapping was subsequently applied to an independent frequency-entangled test state without any additional training or parameter optimization, increasing the Gaussian-shape fidelity between the measured and Gaussian approximation JSIs from 76.2\% to 90\%.

The successful recovery of the independent test state indicates that the reconstructed correction mapping captures the inverse relation needed to compensate for the channel-induced coordinate deformation, rather than features specific to the reference state. This result demonstrates the transferability of the proposed framework and establishes TF grid states not only as engineered quantum states, but also as practical metrological resources for diagnosing and correcting distortions in TF quantum systems.

Although the observed distortion is likely dominated by higher-order dispersion in the present experiment, the proposed approach is fundamentally independent of the specific perturbation mechanism and can, in principle, be extended to arbitrary forms of coordinate deformation in TF distributions. Since the reconstruction procedure does not require prior knowledge of the physical origin of the distortion, it provides a general framework for probing, characterizing, and correcting unknown perturbations in TF quantum systems. 

These capabilities open a pathway toward distortion-resilient TF quantum-state characterization and may find applications in quantum communication, quantum sensing, and high-dimensional quantum information processing.


\section*{Acknowledgments}
This work was supported by the National Science and Technology Council (NSTC) of Taiwan under Grants 112-2112-M-008-025-MY3, 113-2119-M-008-010, 111-2627-M-008-001, 113-2923-E-008-001, and 114-2927-I-008-501. The authors also thank Dr. Bo-Han Wu at University of Hawaii at Mānoa for fruitful discussions on machine learning.

\section*{Appendix}
\appendix

\section{List of Symbols}\label{symbols}

Table \ref{tab:symbols} summarizes the frequently used symbols and their corresponding definitions in this work. It is intended to facilitate the understanding of the data-processing procedures and the relationships among the quantities used throughout the manuscript.

\begin{table}[b]
    \centering
    \caption{Summary of frequently used symbols.}
    \begin{tabular}{|l|l|}
    \hline
     $\Omega_{s,i}$ & Signal and idler photon frequencies \\ \hline
     $t_{s,i}$  & Temporal coordinates \\ \hline
     $t^m_{s,i}$ & Measured temporal coordinates \\ \hline
     $t^{m,g}_{s,i}$ & Measured grid-point coordinates \\ \hline
     $t^{m,g}_{s,0}$,$t^{m,g}_{i,0}$ & Basis grid-point coordinates \\ \hline
     $t^{r}_{s,i}$ &  Reference (ideal) grid-point coordinates \\ \hline
     $F_{s,i}$ & \begin{tabular}{@{}l@{}}Learned correction mapping from measured \\ to reference coordinates \end{tabular} \\ \hline
     $J_{I,uc}^m$, $J_{I,c}^m$ & Uncorrected and corrected measured JSIs \\ \hline
     $J_{I,uc}^G$, $J_{I,c}^G$ & \begin{tabular}{@{}l@{}}Gaussian approximations of the uncorrected\\ and corrected JSIs \end{tabular}\\ \hline
    \end{tabular}
    \label{tab:symbols}
\end{table}

\bibliography{refs}

@article{mower2013high,
  title={High-dimensional quantum key distribution using dispersive optics},
  author={Mower, Jacob and Zhang, Zheshen and Desjardins, Pierre and Lee, Catherine and Shapiro, Jeffrey H and Englund, Dirk},
  journal={Physical Review A—Atomic, Molecular, and Optical Physics},
  volume={87},
  number={6},
  pages={062322},
  year={2013},
  publisher={APS}
}

@article{jin2024spectrally,
  title={Spectrally resolved Franson interference},
  author={Jin, Rui-Bo and Zeng, Zi-Qi and Xu, Dan and Yuan, Chen-Zhi and Li, Bai-Hong and Wang, You and Shimizu, Ryosuke and Takeoka, Masahiro and Fujiwara, Mikio and Sasaki, Masahide and others},
  journal={Science China Physics, Mechanics \& Astronomy},
  volume={67},
  number={5},
  pages={250312},
  year={2024},
  publisher={Springer}
}

@article{fabre2023teleportation,
  title={Teleportation-Based Error Correction Protocol of Time--Frequency Qubit States},
  author={Fabre, Nicolas},
  journal={Applied Sciences},
  volume={13},
  number={16},
  pages={9462},
  year={2023},
  publisher={MDPI}
}

@article{wang2025harnessing,
  title={Harnessing hybrid frequency-entangled qudits through quantum interference},
  author={Wang, Sheng-Hung and Chen, Po-Han and Yang, Cheng-Yu and Chen, Yen-Hung and Tsai, Pin-Ju},
  journal={Physical Review Research},
  volume={7},
  number={4},
  pages={043152},
  year={2025},
  publisher={APS}
}

@article{park2018time,
  title={Time-energy entangled photon pairs from Doppler-broadened atomic ensemble via collective two-photon coherence},
  author={Park, Jiho and Jeong, Taek and Kim, Heonoh and Moon, Han Seb},
  journal={Physical Review Letters},
  volume={121},
  number={26},
  pages={263601},
  year={2018},
  publisher={APS}
}

@article{jin2016simple,
  title={Simple method of generating and distributing frequency-entangled qudits},
  author={Jin, Rui-Bo and Shimizu, Ryosuke and Fujiwara, Mikio and Takeoka, Masahiro and Wakabayashi, Ryota and Yamashita, Taro and Miki, Shigehito and Terai, Hirotaka and Gerrits, Thomas and Sasaki, Masahide},
  journal={Quantum Science and Technology},
  volume={1},
  number={1},
  pages={015004},
  year={2016},
  publisher={IOP Publishing}
}

@article{fabre2020generation,
  title={Generation of a time-frequency grid state with integrated biphoton frequency combs},
  author={Fabre, Nicolas and Maltese, Giorgio and Appas, F{\'e}licien and Felicetti, Simone and Ketterer, Andreas and Keller, Arne and Coudreau, Thomas and Baboux, Florent and Amanti, Maria I and Ducci, S and others},
  journal={Physical Review A},
  volume={102},
  number={1},
  pages={012607},
  year={2020},
  publisher={APS}
}

@article{descamps2024gottesman,
  title={Gottesman-Kitaev-Preskill encoding in continuous modal variables of single photons},
  author={Descamps, {\'E}loi and Keller, Arne and Milman, P{\'e}rola},
  journal={Physical Review Letters},
  volume={132},
  number={17},
  pages={170601},
  year={2024},
  publisher={APS}
}

@article{brecht2015photon,
  title={Photon temporal modes: a complete framework for quantum information science},
  author={Brecht, Benjamin and Reddy, Dileep V and Silberhorn, Christine and Raymer, Michael G},
  journal={Physical Review X},
  volume={5},
  number={4},
  pages={041017},
  year={2015},
  publisher={APS}
}

@article{gianani2020measuring,
  title={Measuring the time--frequency properties of photon pairs: A short review},
  author={Gianani, Ilaria and Sbroscia, Marco and Barbieri, Marco},
  journal={AVS Quantum Science},
  volume={2},
  number={1},
  year={2020},
  publisher={AIP Publishing}
}

@article{chang2025recent,
  title={Recent advances in high-dimensional quantum frequency combs},
  author={Chang, Kai-Chi and Cheng, Xiang and Sarihan, Murat Can and Wong, Chee Wei},
  journal={Newton},
  volume={1},
  number={1},
  year={2025},
  publisher={Elsevier}
}

@article{yu2024time,
  title={Time-encoded photonic quantum states: Generation, processing, and applications},
  author={Yu, Hao and Govorov, Alexander O and Song, Hai-Zhi and Wang, Zhiming},
  journal={Applied Physics Reviews},
  volume={11},
  number={4},
  year={2024},
  publisher={AIP Publishing}
}

@article{pousset2026time,
  title={Time-frequency Talbot effect as Clifford operations on entangled time-frequency GKP states},
  author={Pousset, Thomas and Dalidet, Romain and Labont{\~A}{\v{S}}, Laurent and Fabre, Nicolas},
  journal={arXiv preprint arXiv:2603.24279},
  year={2026}
}

@article{liu2019energy,
  title={Energy-time entanglement-based dispersive optics quantum key distribution over optical fibers of 20 km},
  author={Liu, Xu and Yao, Xin and Wang, Heqing and Li, Hao and Wang, Zhen and You, Lixing and Huang, Yidong and Zhang, Wei},
  journal={Applied Physics Letters},
  volume={114},
  number={14},
  year={2019},
  publisher={AIP Publishing}
}

@article{chang2024large,
  title={Large-alphabet time-bin quantum key distribution and Einstein--Podolsky--Rosen steering via dispersive optics},
  author={Chang, Kai-Chi and Sarihan, Murat Can and Cheng, Xiang and Zhang, Zheshen and Wong, Chee Wei},
  journal={Quantum Science and Technology},
  volume={9},
  number={1},
  pages={015018},
  year={2024},
  publisher={IOP Publishing}
}

@article{liu2024high,
  title={High-dimensional quantum key distribution using energy-time entanglement over 242 km partially deployed fiber},
  author={Liu, Jingyuan and Lin, Zhihao and Liu, Dongning and Feng, Xue and Liu, Fang and Cui, Kaiyu and Huang, Yidong and Zhang, Wei},
  journal={Quantum Science and Technology},
  volume={9},
  number={1},
  pages={015003},
  year={2024},
  publisher={IOP Publishing}
}

@article{lu2023frequency,
  title={Frequency-bin photonic quantum information},
  author={Lu, Hsuan-Hao and Liscidini, Marco and Gaeta, Alexander L and Weiner, Andrew M and Lukens, Joseph M},
  journal={Optica},
  volume={10},
  number={12},
  pages={1655--1671},
  year={2023},
  publisher={Optica Publishing Group}
}

@article{davis2017pulsed,
  title={Pulsed single-photon spectrometer by frequency-to-time mapping using chirped fiber Bragg gratings},
  author={Davis, Alex OC and Saulnier, Paul M and Karpi{\'n}ski, Micha{\l} and Smith, Brian J},
  journal={Optics Express},
  volume={25},
  number={11},
  pages={12804--12811},
  year={2017},
  publisher={Optical Society of America}
}

@book{williams2006gaussian,
  title={Gaussian processes for machine learning},
  author={Williams, Christopher KI and Rasmussen, Carl Edward},
  volume={2},
  number={3},
  year={2006},
  publisher={MIT press Cambridge, MA}
}

@article{chang2026gkp,
  title={GKP-inspired high-dimensional superdense coding with energy-time entanglement},
  author={Chang, Kai-Chi and Mirani, Arjun and Sarihan, Murat Can and Cheng, Xiang and Harasimowicz, Michelle and Hayden, Patrick and Wong, Chee Wei},
  journal={arXiv preprint arXiv:2602.15125},
  year={2026}
}

@article{maclean2018direct,
  title={Direct characterization of ultrafast energy-time entangled photon pairs},
  author={MacLean, Jean-Philippe W and Donohue, John M and Resch, Kevin J},
  journal={Physical review letters},
  volume={120},
  number={5},
  pages={053601},
  year={2018},
  publisher={APS}
}

@article{graffitti2020direct,
  title={Direct generation of tailored pulse-mode entanglement},
  author={Graffitti, Francesco and Barrow, Peter and Pickston, Alexander and Bra{\'n}czyk, Agata M and Fedrizzi, Alessandro},
  journal={Physical Review Letters},
  volume={124},
  number={5},
  pages={053603},
  year={2020},
  publisher={APS}
}

@article{merkouche2022heralding,
  title={Heralding multiple photonic pulsed bell pairs via frequency-resolved entanglement swapping},
  author={Merkouche, Sofiane and Thiel, Val{\'e}rian and Davis, Alex OC and Smith, Brian J},
  journal={Physical Review Letters},
  volume={128},
  number={6},
  pages={063602},
  year={2022},
  publisher={APS}
}

@article{merkouche2022spectrally,
  title={Spectrally resolved four-photon interference of time-frequency-entangled photons},
  author={Merkouche, Sofiane and Thiel, Val{\'e}rian and Smith, Brian J},
  journal={Physical Review A},
  volume={105},
  number={2},
  pages={023708},
  year={2022},
  publisher={APS}
}

@article{fabre2025photonic,
  title={Photonic quantum information processing using the frequency continuous variable of single photons: N. Fabre, U. Chabaud},
  author={Fabre, Nicolas and Chabaud, Ulysse},
  journal={Quantum Information Processing},
  volume={25},
  number={1},
  pages={7},
  year={2025},
  publisher={Springer}
}

@article{gayer2008temperature,
  title={Temperature and wavelength dependent refractive index equations for MgO-doped congruent and stoichiometric LiNbO3},
  author={Gayer, O and Sacks, Z and Galun, E and Arie, A},
  journal={Applied Physics B},
  volume={91},
  number={2},
  pages={343--348},
  year={2008},
  publisher={Springer}
}

@article{maclean2019reconstructing,
  title={Reconstructing ultrafast energy-time-entangled two-photon pulses},
  author={MacLean, Jean-Philippe W and Schwarz, Sacha and Resch, Kevin J},
  journal={Physical Review A},
  volume={100},
  number={3},
  pages={033834},
  year={2019},
  publisher={APS}
}

@article{fabre2022time,
  title={Time and frequency as quantum continuous variables},
  author={Fabre, Nicolas and Keller, Arne and Milman, P{\'e}rola},
  journal={Physical Review A},
  volume={105},
  number={5},
  pages={052429},
  year={2022},
  publisher={APS}
}

@article{yoon2026hardware,
  title={A hardware-native time-frequency GKP logical qubit toward fault-tolerant photonic operation},
  author={Yoon, Tai Hyun},
  journal={arXiv preprint arXiv:2602.14461},
  year={2026}
}
\end{document}